\documentclass[letterpaper]{article} 
\usepackage{aaai2026}  
\usepackage{times}  
\usepackage{helvet}  
\usepackage{courier}  
\usepackage[hyphens]{url}  
\usepackage{graphicx} 
\usepackage{natbib}  
\usepackage{caption} 
\usepackage{algorithm}
\usepackage{algorithmic}
\usepackage{amsmath}
 \usepackage{amssymb} 
\usepackage{color}
\usepackage{booktabs}
\usepackage{multirow}
\usepackage{newfloat}
\usepackage{listings}
\DeclareCaptionStyle{ruled}{labelfont=normalfont,labelsep=colon,strut=off} 
\floatstyle{ruled}
\newfloat{listing}{tb}{lst}{}
\floatname{listing}{Listing}
\title{CLIPPan: Adapting CLIP as A Supervisor for Unsupervised Pansharpening}
\author{
    Lihua Jian\textsuperscript{\rm 1}, 
    Jiabo Liu\textsuperscript{\rm 1},
    Wushao Wu\textsuperscript{\rm 2},
    Lihui Chen\textsuperscript{\rm 3}\thanks{Corresponding author.}\\
}
\affiliations{
    \textsuperscript{\rm 1}School of Electrical and Information Engineering, Zhengzhou University, China\\
    \textsuperscript{\rm 2}School of computer science, Wuhan University, China\\
    \textsuperscript{\rm 3}School of Microelectronics and Communication Engineering, Chongqing University, China \\

    lihui.chen@cqu.edu.cn
}

\begin{document}

\maketitle

\begin{abstract}
Despite remarkable advancements in supervised pansharpening neural networks, these methods face domain adaptation challenges of resolution due to the intrinsic disparity between simulated reduced-resolution training data and real-world full-resolution scenarios.To bridge this gap, we propose an unsupervised pansharpening framework, CLIPPan, that enables model training at full resolution directly by taking CLIP, a visual-language model, as a supervisor. However, directly applying CLIP to supervise pansharpening remains challenging due to its inherent bias toward natural images and limited understanding of pansharpening tasks. 
Therefore, we first introduce a lightweight fine-tuning pipeline that adapts CLIP to recognize low-resolution multispectral, panchromatic, and high-resolution multispectral images, as well as to understand the pansharpening process. Then, building on the adapted CLIP, we formulate a novel \textit{loss integrating semantic language constraints}, which aligns image-level fusion transitions with protocol-aligned textual prompts (e.g., Wald's or Khan's descriptions), thus enabling CLIPPan to use language as a powerful supervisory signal and guide fusion learning without ground truth. Extensive experiments demonstrate that CLIPPan consistently improves spectral and spatial fidelity across various pansharpening backbones on real-world datasets, setting a new state of the art for unsupervised full-resolution pansharpening.
\end{abstract}

%
\begin{links}
    \link{Code}{https://github.com/Jiabo-Liu/CLIPPan}
\end{links}

\section{Introduction}

{Pansharpening fuses multispectral (MS) images with rich spectral information and panchromatic (PAN) images with detailed spatial information to achieve high-resolution MS (HRMS) images, showing significant potential in remote sensing applications, such as urban planning and environmental surveillance \cite{a1}. }

\begin{figure}
\centering
\includegraphics[trim=5mm 5mm 5mm 5mm, width=\linewidth]{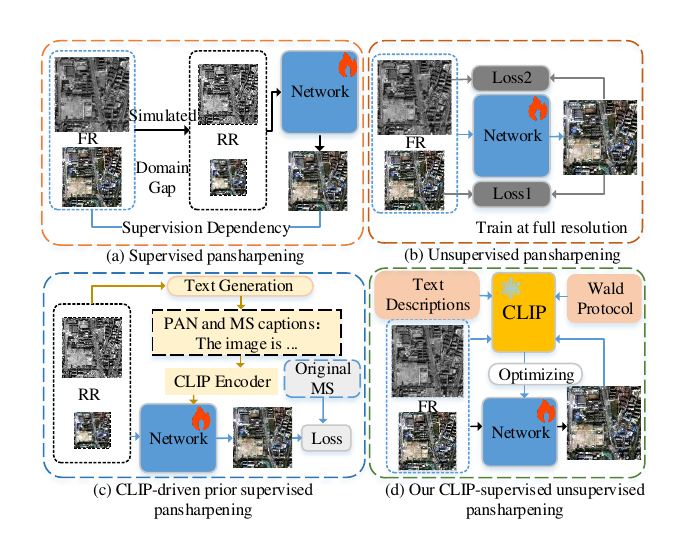}
\caption{The comparison of the different pansharpening paradigms.}
\label{fig:intro}
\end{figure}

{Over the years, lots of methods have been developed for pansharpening, including component substitution (CS) \cite{ref6,ref7,ref9}, multi-resolution analysis (MRA) \cite{ref11,ref12,ref13}, variational optimization (VO) \cite{ref15,ref16}, and deep learning (DL) -based approaches \cite{ref17,deng2025}. Nevertheless, the former three types of methods usually struggle to balance spectral and spatial fidelity or rely on hand-crafted priors. In contrast, DL-based methods not only avoid these limitations but also dominate this topic recently for their superior ability to learn complex mappings \cite{ref18}.} 

However, as shown in Figure~1(a), most DL-based methods rely on ground truth (GT) for supervision of the models, such as the classical PNN \cite{ref17}, PanNet \cite{ref18}, and SDRCNN \cite{ref19}. Despite the continuous proposal of various methods, most of them focus on architecture designing \cite{tianxin} or novel regularization terms \cite{ref20} instead of eliminating the dependency on GT. 
Unfortunately, the GT is inaccessible for MS images in the real scenario at full resolution. To this end, supervised methods are usually trained by simulated data of LRMS, PAN, and HRMS images at reduced resolution, leading to a significant performance degradation on real full-resolution imagery due to domain gaps of scale.  

Therefore, unsupervised methods that directly train models at full resolution have attracted the attention of researchers for pansharpening. To avoid the reliance on GT, most unsupervised methods generally train models by employing GAN-based architectures \cite{ref21}, priors \cite{uns-prior}, or spectral-spatial consistency \cite{spe-spa} between the fused and source images. However, due to the absence of the GT, these unsupervised methods (Figure~\ref{fig:intro} (b)) lack direct guidance to learn pansharpening. Besides, these methods only impose the regularization \cite{uns1,uns2,uns3} by a low-level relationship between the fused output and the source images, {thus hardly ensuring the output in the HRMS domain.}

To address these challenges, a potential solution is to tell the pansharpening model the fusion objective or rule (e.g., Wald's or Khan's protocols). In this way, high-level semantic supervision with texture prompts can be utilized to constrain the fused output in the HRMS domain when GT is inaccessible. Therefore, inspired by the remarkable capability of aligning images with texture prompts in a shared semantic space of recent vision-language models, particularly CLIP, we explore an unsupervised pansharpening framework at full resolution by providing textual prompts (e.g., Wald’s protocol) as supervisory signals for pansharpening models in the absence of GT. 

However, despite its superior alignment capability, CLIP's strong bias towards natural images and lack of understanding for pansharpening hinder it a competent supervisor to guide the pansharpening model. Moreover, trained by RGB images, CLIP cannot tackle MS images with more bands and recognize the spectral characters of MS images.
Therefore, we first adapt the pre-trained CLIP model to a qualified supervisor by \textit{i}) establishing reliable modality recognition by binding LRMS, PAN, and HRMS images to their corresponding semantic spaces represented by textual prompts through inter-modal contrastive learning (Inter-MCL); \textit{ii}) enabling CLIP to recognize image content of remote sensing images and maintain feature diversity via intra-modal contrastive learning (IntraMCL); and \textit{iii}) guiding CLIP to understand the pansharpening by aligning fused image features with texture prompts of fusion protocols such as Wald's rule. Once adapted, CLIP serves as a fixed semantic supervisor to guide the pansharpening network via a joint loss integrating low-level visual and semantic texture constraints, effectively bridging the fused outputs with the HRMS domain without requiring any GT labels.

In conclusion, the contributions are as follows.
\begin{itemize}
    \item 
    We present CLIPPan, a universal framework to leverage vision-language models (particularly CLIP) for unsupervised full-resolution pansharpening via protocol-informed linguistic guidance. Compatible with any pansharpening backbone, CLIPPan achieves state-of-the-art performance on various datasets, significantly enhancing both spectral and spatial fidelity in real-world scenarios.
    \item 
    We design a lightweight CLIP adaptation strategy tailored for pansharpening, enhancing CLIP's ability to recognize remote sensing images and the pansharpening process, paving the way for future research in unsupervised pansharpening based on visual-language models.
    \item 
    We introduce a novel language-guided unsupervised loss based on Wald's protocols, which provides semantic alignment between the fused output and the HRMS domain, contributing to future works in utilizing textual prompts to supervise the pansharpening. Besides, this research can establish a reciprocal framework. That is, the underlying paradigm can conversely evaluate the effectiveness of protocols and even guide the discovery of novel pansharpening protocols.
\end{itemize}

\section{Related Works}
\begin{figure*}[t!]
\centering 
\includegraphics[width=0.9 \textwidth]{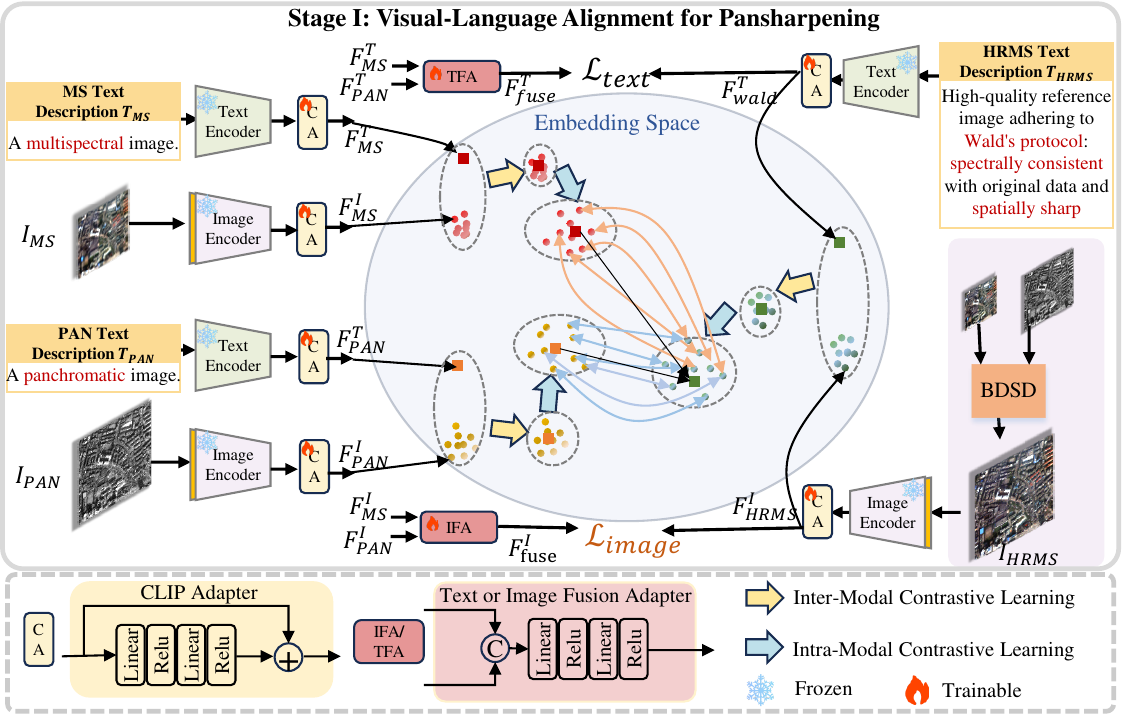}
\caption{Workflow for stage I. Visual-language alignment for pansharpening.}
\label{fig_2}
\end{figure*}

\subsection{DL-Based Pansharpening}

Optimized by pixel-level losses (e.g., $\ell_1$ or $\ell_2$) between the fused result and the ground truth, supervised methods are trained at reduced resolution using paired triplets of LRMS, PAN, and HRMS images. Representative works include PNN \cite{ref17}, PanNet \cite{ref18}, and SDRCNN \cite{ref19} as well as more recent architectures like ADKNet \cite{ref25}, PSCINN \cite{wang2024pan}, SSFMamba \cite{ssfmamba}, and FusionMamba \cite{fusionmamba}. Despite strong performance on synthetic datasets, these methods suffer from poor generalization to real full-resolution images due to the domain gap between the simulated reduced-resolution and real full-resolution data. Although a recent study \cite{deng2025} shown in Figure~\ref{fig:intro} (c) attempted to introduce CLIP-derived priors, it still dependent on low-resolution settings and handcrafted protocols, leaving the full-resolution unsupervised setting underexplored. Differently, our CLIPPan shown in Figure~\ref{fig:intro} (d) allows direct training at full resolution with both semantic and low-level visual constraints, eliminating reliance on GT while improving spectral-spatial fidelity.

{Unsupervised methods avoid using HRMS labels and instead design training objectives based on low-level relationships between the fused image and its inputs, such as spatial-spectral consistency \cite{spe-spa}, modality alignment \cite{uns-mod}, or detail preservation \cite{uns-deta}. Some approaches adopt handcrafted priors \cite{ref20}, while others use generative frameworks like GANs \cite{ref21,zhu2023qis,ssa-gan} or diffusion models \cite{difuss1,frdiff} to better model fusion distributions. However, these methods still rely on heuristic, low-level visual constraints that offer limited guidance on what constitutes a semantically valid or perceptually high-quality fusion.}

\subsection{Visual-Language Alignment}
While vision–language models like CLIP \cite{clip} have demonstrated powerful generalization across modalities and domains, directly applying them to remote-sensing tasks remains non-trivial. This is due to both the modality gap—satellite imagery significantly differs from natural images in texture, geometry, and semantics—and the task gap, as standard CLIP is trained for image-text alignment rather than pansharpening.

To adapt CLIP for downstream tasks, numerous parameter-efficient fine-tuning strategies have been proposed. CLIP-Adapter \cite{clip-ada} inserts lightweight bottleneck layers to inject task-specific knowledge; prompt-based tuning methods such as CoOp \cite{coop} and CoCoOp \cite{cocoop} learn input-dependent textual prompts to steer the joint embedding space; LoRA-CLIP \cite{clip-lora} applies low-rank adaptation to reduce training overhead. In remote sensing, recent methods like RS-CLIP \cite{rsclip} and GeoCLIP \cite{geoclip} incorporate domain-specific prompts or geospatial priors to bridge the semantic gap between satellite images and natural images. 

Unlike prior works that focus on classification or segmentation, we take a novel step by adapting CLIP as a semantic supervisor for an unsupervised pansharpening task.

\section{Methodology}
\begin{figure}[!t]
\centering 
\includegraphics[width=0.45 \textwidth]{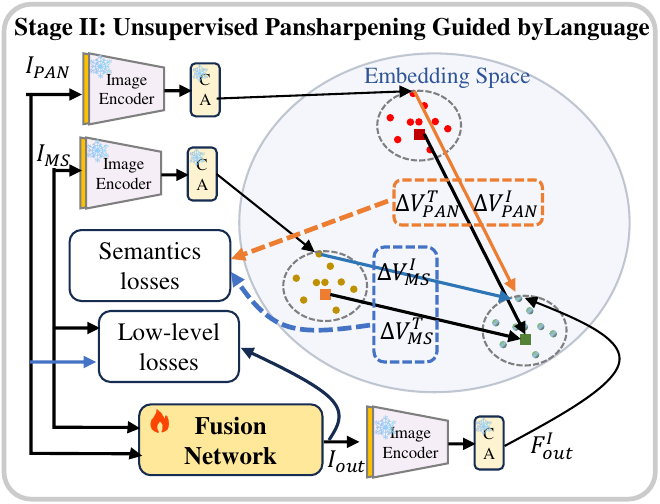}
\caption{Workflow for stage II. Unsupervised pansharpening guided by language.}
\label{fig:stage2}
\end{figure}

As shown in Figure~\ref{fig_2} and Figure~\ref{fig:stage2}, the proposed CLIPPan achieves unsupervised pansharpening by two stages. At the first stage, CLIP is adapted to align the LRMS, PAN, and HRMS images with the corresponding semantic space described by texture prompts, respectively. Notably, the HRMS images are particularly aligned to a semantic space described by the panshaprening objective, e.g., Wald's protocol. Then, at the second stage, the adapted CLIP supervises the pansharpening model by evaluating whether the fused out align with the pansharpening objective or not, thus eliminating the dependence on the GT.

\subsection{Stage I: Visual-Language~Alignment~for~Pansharpening}
Despite superior alignment performance on web images and language, the pretrained CLIP face challenges of bias toward natural images and incompatibility with MS image bands for the purpose. Therefore, adapting CLIP is critical for pansharpening. Meanwhile, to maintain the strong generalization of CLIP, a parameter-efficient fine-tuning stratege is adopted in our adaptation. Specifically, we introduce six lightweight adapter modules (i.e., CA modules in Figure~\ref{fig_2}):  three after the visual encoder for vision adaptation, and the other three after the text encoder for text adaptation. Besides, since the \textbf{visual} encoder is incompatible with the MS images, we replace the original layer designed for natural RGB images with a convolutional layer specific for MS inputs.

To make the CLIP a proficient supervisor for pansharpening, the adaptation of is optimized by the following loss:
\begin{equation}
  \mathcal{L}_{\text{s1}} = \mathcal{L}_{\text{inter}} + \mathcal{L}_{\text{intra}} +\mathcal{L}_{\text{fusion}}, 
\end{equation}
where $\mathcal{L}_{\text{intra}}$, the InterMCL loss, is used to bind image {types} of LRMS, PAN, and HRMS to their semantic spaces, respectively; $ \mathcal{L}_{\text{inter}}$, the IntraMCL loss, is used to ensure the adapted CLIP to recognize image content and feature diversity; and $\mathcal{L}_{\text{fusion}}$, the fusion loss, is employed to ensure that the feature representations of MS and PAN images can be projected into the HRMS feature space.

\subsubsection{Inter-Modal Contrastive Learning (InterMCL).}
To adapt CLIP to remote sensing images, an intuitive attempt is to utilize the vanilla language-image contrastive learning in CLIP. However, for finetuning a pansharpening supervisor, the vanilla way is inappropriate to adapt CLIP. Since, as a pansharpening supervisor, the model needs the ability to discriminate the MS, PAN, and HRMS images, thereby utilizing its understanding of image types to supervise the learning of pansharpening models. Nevertheless, the vanilla contrastive language-image learning utilizes content-dependent descriptions for various images, aiming at recognizing image content by language prompts, which is inconsistent with the objective of recognizing different image types. Besides, it is difficult to create a dataset with quadruples of LRMS-PAN-HRMS-text. 

 Therefore, to bind an image type to a semantic space, a better choice is to use semantically similar descriptions for a distinct image type instead of content-dependent descriptions for various images. For simple, constant descriptions for distinct types of images are used in the proposed method texture, i.e., $F_\text{MS}=$\textit{``a multispectral image''}, $T_\text{PAN}=$\textit{``a panchromatic image''}, $T_\text{HRMS}=$\textit{``High-quality reference image adhering to Wald's protocol: spectrally consistent with original data and spatially sharp''}. Meanwhile, considering that HRMS images cannot be obtained at full resolution in real cases, HRMS images are generated on-the-fly using the conventional BDSD \cite{ref44} algorithm.

 Specifically, as shown in Figure~\ref{fig_2}, we pair every image in a specific type, e.g., $F_\text{MS}^\text{I}$, with its corresponding text prompt (i.e., $F_\text{MS}^T$). Then, these matched pairs are treated as positives, while all other image–text combinations within the batch serve as negatives. This encourages the model to pull together the paired image-text samples {semantically aligned representations} while pushing apart mismatched ones, thus enabling CLIP to recognize image types and binding image types to the corresponding semantic spaces. In detail, the inter-MCL can be formulated as follows:
\begin{equation}
\begin{aligned}
\mathcal{L}_{\text{inter}} &= \frac{1}{3} \sum_{M1,M2} \mathcal{L}_{\text{align}}(F^I_{M1}, F^T_{M2}),\\
\mathcal{L}_{\text{align}} &= -\frac{1}{N}\sum_{i=1}^N \log \frac{\exp(\langle F^{I(i)}_{M1}, F^{T(i)}_{M2} \rangle / \tau_c)}{\sum_{j=1}^N \exp(\langle F^{I(i)}_{M1}, F^{T(j)}_{M2} \rangle / \tau_c)},
\label{eq:inter}
\end{aligned}
\end{equation}
where $M1, M2\in \{ \text{MS}, \text{PAN}, \text{HRMS}\}$; $F^{I(i)}_{M1}$ and $F^{T(i)}_{M2}$ are the $i$-th adapted $M1$-type image and $M2$-{type} text features, $\langle \mathbf{a}, \mathbf{b} \rangle = \mathbf{a}^\top \mathbf{b} / (\|\mathbf{a}\|\|\mathbf{b}\|)$ computes the cosine similarity between vectors, and $\tau$ is the temperature controlling alignment strength. This loss ensures that the adapted CLIP preserves language-image alignment necessary for downstream semantic supervision.


\subsubsection{Intra-Modal Contrastive Learning (IntraMCL).}
However, using constant descriptions for all images of a specific image type to adapt CLIP easily results in feature collapsing to the fixed embedding representation of its corresponding texture {prompt}. Besides, it decreases CLIP's capability of recognizing image content. 

Therefore, we additionally introduce contrastive learning within the image domain by taking the same scene of LRMS, PAN, and HRMS images as positive samples, while others as negative samples. Therefore, images with similar geographic scenes are closing while images with different scenes are diverging, thus ensuring diversity of image features and discriminating different image semantic content. Besides, IntraMCL facilitates domain transfer from natural imagery to remote sensing and bridges the gap from natural to remote sensing domain.

Specifically, we construct training batches by sampling $N$ paired triplets consisting of LRMS, PAN, and HRMS patches. 
Then, the IntraMCL can be formulated as:


\begin{align}
\mathcal{L}_{\text{intra}} = -\frac{1}{3N} \sum_{i=1}^{3N} \log \frac{\exp(\langle F^{I(i)}_{M1}, F^{I(i)}_{M2} \rangle / \tau_i)}{\sum_{k=1}^{3N} \exp(\langle F^{I(i)}_{M1}, F^{I(j)}_{M1} \rangle / \tau_i)}.
\label{eq:intra}
\end{align}


\subsubsection{Fusion-Aware Alignment.}

Although InterMCL and IntraMCL bind various image types to a specific semantic space while maintaining feature diversity, it does not explicitly model the fusion between MS and PAN inputs, lacking understanding of fusion processing. Therefore, we enforce the fusion learning in the adaptation of CLIP, 
so that CLIP can recognize the fused HRMS representation as a meaningful combination of its sources.

{To this end, we introduce two auxiliary modules, i.e., the image fusion adapter (IFA) and the text fusion adapter (TFA) in Figure~\ref{fig_2}. These adapters operate on encoded features and learn to generate fused image and text embeddings from LRMS and PAN inputs:}
\begin{equation}
    F^I_{\mathrm{fuse}}=\mathrm{IFA}(F^I_{\mathrm{MS}}, F^I_{\mathrm{PAN}}), F^T_{\mathrm{fuse}}=\mathrm{TFA}(F^T_{\mathrm{MS}}, F^T_{\mathrm{PAN}}),
\end{equation}
By enforcing alignment between these fused image/text features and the corresponding reference features, i.e., the (HRMS image)/(Wald's protocol text) features, we guide the model to internalize the mapping from source modalities to a semantically valid fusion target. In detail, the alignment is achieved through simple $\mathcal{L}_1$ loss: 
\begin{equation}
\mathcal{L}_{\text{fusion}} = \|F^T_{\text{fuse}} - F^T_{\text{wald}}\|_1 + \|F^I_{\text{fuse}} - F^I_{\text{HRMS}}\|_1.
\end{equation}
Eventually, the adapted CLIP acquires the ability to project MS and PAN image features into the semantic space of high-quality HRMS images, which facilitates better supervision for the pansharpening model in Stage II.

\subsection{Stage II: Unsupervised~Pansharpening~Guided~by~Language}

To achieve unsupervised pansharpening, we combine the semantic and low-level supervision for the pansharpening model and leverage their complementary advantages.


\subsubsection{Semantic Supervision by Language.}
  
Thanks to the adaptation in Stage I, the fusion objective described in the textual prompt (i.e., Wald's protocol) is aligned to the domain of HRMS images. Therefore, the adapted CLIP can naturally evaluate the quality of fusion by judging if the fused output is aligned with the semantic features extracted from the Wald's protocol or not. However, we cannot use element-wise loss between the visual features extracted from the adapted CLIP with features of Wald's protocol, due to the latter's invariance for all fused images.

Nevertheless, as show in Figure~\ref{fig:stage2}, the vectors, i.e., $ \Delta \mathbf{V}^T_{\text{MS}}$, from features of $F_\text{MS}^\text{T}$ to Wald's $F_\text{wald}^\text{T}$ can reflect the feature transition of fusion. Therefore, we employ a directional vector from source to target text to guide the training of the pansharpening network. Specifically, the pansharpening network can be supervised by minimizing the angular discrepancy between the feature displacement vectors of image pairs and their corresponding text pairs, i.e.
\begin{equation}
\mathcal{L}_d = 1 - \frac{1}{2} \left( 
\langle \Delta \mathbf{V}^I_{\text{MS}}, \Delta \mathbf{V}^T_{\text{MS}} \rangle + 
\langle \Delta \mathbf{V}^I_{\text{PAN}}, \Delta \mathbf{V}^T_{\text{PAN}} \rangle \right), \label{eq:semantic-loss}
\end{equation}
\begin{align}
\Delta \mathbf{V}^I_{\text{MS}} &= F^I_{\text{out}} - F^I_{\text{MS}}, &
\Delta \mathbf{V}^T_{\text{MS}} &= F^T_{\text{wald}} - F^T_{\text{MS}}, \\
\Delta \mathbf{V}^I_{\text{PAN}} &= F^I_{\text{out}} - F^I_{\text{PAN}}, &
\Delta \mathbf{V}^T_{\text{PAN}} &= F^T_{\text{wald}} - F^T_{\text{PAN}}.
\end{align}
where $F^I_{\text{out}}$ denotes the fused image embedding, and $F^I_{\text{MS}}$, $F^I_{\text{PAN}}$ be the embeddings of MS and PAN inputs, respectively. $F^T_{\text{wald}}$, $F^T_{\text{MS}}$, and $F^T_{\text{PAN}}$ are the textual embeddings, repsectively.

Consequently, through penalizing angular misalignment between transitions in the image and text shared embedding spaces, the pansharpening model is encouraged to produce outputs semantically aligned with the HRMS image domain. 


\subsubsection{Low-Level Unsupervised Reconstruction Losses.}
However, since the semantic supervision by language can only impose the output in the HRMS image domain, the image content and spectral characters are not supervised. Therefore, we also introduce the low-level visual constraints for unsupervised pansharpening. 

Specifically, the low-level visual unsupervised loss function $\mathcal{L}_{\text{unsup}}$ includes three key components to simultaneously enforce spectral fidelity, spatial sharpness, and perceptual quality. The spectral fidelity is achieved by 
\begin{equation}
\begin{aligned}
\mathcal{L}_{\text{spec}} &= \|\downarrow(\mathbf{I}_{\text{out}}) - \mathbf{I}_{\text{MS}}\|_2^2 +1 - \text{SSIM}\left(\downarrow(\mathbf{I}_{\text{out}}), \mathbf{I}_{\text{MS}}\right),
\end{aligned}
\end{equation}
where $\downarrow$ denotes bicubic downsampling (4× ratio), $\mathbf{I}_{\text{out}}$ is the fused HRMS output and $\text{SSIM}$ \cite{ssim} measures structural similarity at reduced resolution. The spatial sharpness is achieved by:
\begin{equation}
\begin{aligned}
\mathcal{L}_{\text{spat}} &= \|\phi(\mathbf{I}_{\text{out}}) - \mathbf{I}_{\text{PAN}} \|_2^2 + 1 - \text{SSIM}(\phi(\mathbf{I}_{\text{out}}), \mathbf{I}_{\text{PAN}}), \\
\end{aligned}
\end{equation}
where $\phi(\cdot)$ denotes a {1 × 1 convolution} that degrades the multispectral channels into a single band.
A trade-off loss between spectral and spatial information based on QNR \cite{ref43} is also adopted 
\begin{equation}
\mathcal{L}_\text{QNR}  = (1 - D_\lambda)(1 - D_s).
\end{equation}

Finally, to stabilize the training process, we introduce a pseudo-supervision $\mathcal{L}_{\text{ship}}$ that takes as the reference the output of an existing pansharpening network [SHIP \cite{zhou2024probing} used in this paper] trained at reduced resolution. Therefore, the overall low-level visual loss is
\begin{equation}
\begin{aligned}
\mathcal{L}_{\text{s2}} &= \mathcal{L}_{\text{spec}} + \mathcal{L}_{\text{spat}} + \mathcal{L}_\text{QNR} + \mathcal{L}_{\text{ship}}. \label{eq:low-level-loss}
\end{aligned}
\end{equation}

\begin{figure}[t]
\centering
\includegraphics[trim=8mm 8mm 8mm 8mm, width=0.88\linewidth]{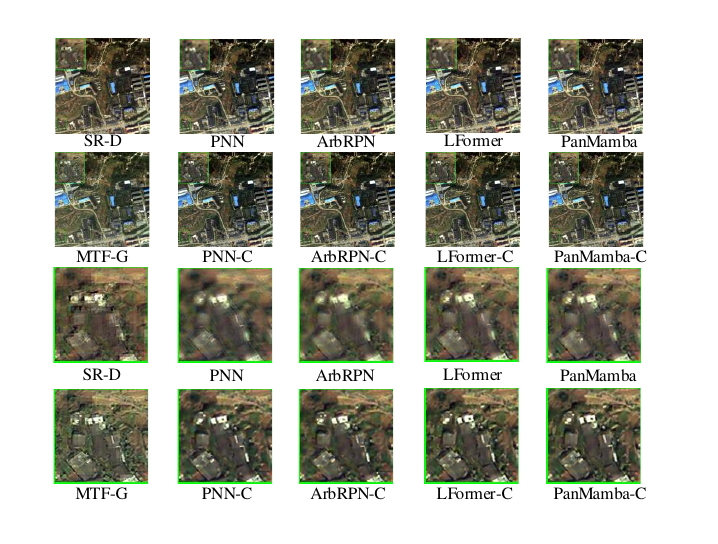}
\caption{Qualitative results of different methods on the full-resolution QB dataset. The top-row images are shown in RGB, and the areas enclosed by the green box have been magnified three times.}
\label{fig_3}
\end{figure}

\begin{figure}[t]
\centering
\includegraphics[trim=8mm 8mm 8mm 8mm, width=0.88\linewidth]{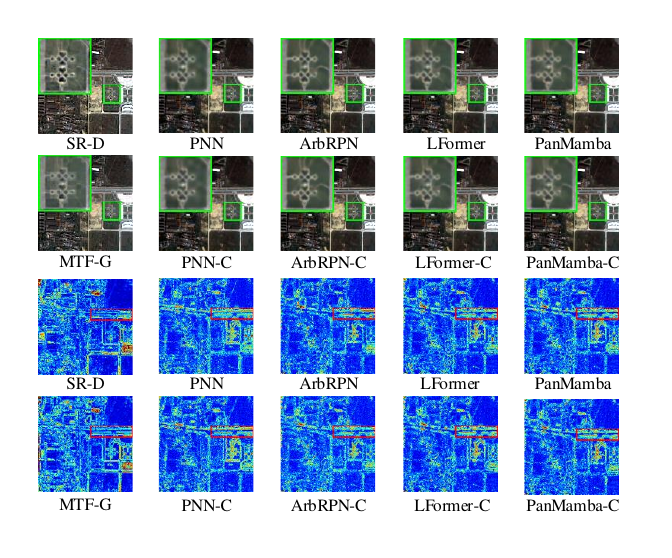}
\caption{Qualitative results of different methods on the reduced-resolution QB dataset. The top-row images are shown in RGB, and the areas enclosed by the green box have been magnified three times. The bottom-row images are error maps.}
\label{fig_4}
\end{figure}

\begin{table*}[t]
\centering
\small
\setlength{\tabcolsep}{2pt} 


\begin{tabular}{cccc|cccc|ccc|cccc}
\toprule
\multirow{2}{*}{Methods} & \multicolumn{7}{c}{QB}  & \multicolumn{7}{c}{WV3} \\
\cmidrule{2-15}
    &$D_{\mathrm{\lambda}}$↓ &$D_{\mathrm{s}}$↓ &QNR↑ &MPSNR↑  &ERGAS↓ &SAM↓ &Q2n↑  &$D_{\mathrm{\lambda}}$↓ &$D_{\mathrm{s}}$↓ &QNR↑  &MPSNR↑  &ERGAS↓ &SAM↓ &Q2n↑ \\
\hline
SR-D  &0.0151 &0.0410 &0.9445 &39.0381 &3.1441 &3.2672 &0.7488 &0.0258 &0.0580 &0.9177 &29.3070 &8.9904 &9.6229 &0.7334 \\
MTF-G &0.0386 &0.0557 &0.9104 &44.3701 &1.8564 &1.2384 &0.8088 &0.0585 &0.0552 &0.8895 &30.5174 &6.2900 &6.4423 &0.7710  \\
\hline
$\lambda$-PNN &0.0604 &0.1003 &0.8467 &42.1526 &1.7170 &1.5809 &0.7263 &0.0722 &0.0565 &0.8757 &31.1314 &6.5617 &6.2032 &0.7006 \\
$\lambda$-PNN-C &0.0063 &0.0294 &0.9643 &44.3535 &1.3345 &1.6468 &0.7858 &0.0091 &0.0364 &0.9547 &33.9501 &4.8606 &5.9487 &0.7914\\
\hline
PNN &0.0137 &0.0347 &0.9521 &48.8494 &0.7372 &0.9404 &0.8799   &0.0245 &0.0358 &0.9405 &37.1830 &3.2368 &4.3787 &0.8324\\
PNN-C  &0.0138 &0.0336 &0.9532 &48.9031 &0.7226 &0.9275 &0.8806 &0.0098 &0.0404 &0.9501 &37.2778 &3.2802 &4.3172 &0.8355 \\
ArbRPN  &0.0140 &0.0281 &0.9582 &51.2758 &0.5470 &0.7299 &0.9079 &0.0271 &0.0356 &0.9383 &38.3459 &2.8467 &3.7834 &0.8538  \\
ArbRPN-C &0.0030 &0.0279 &0.9691 &51.3047 &0.5448 &0.7250 &0.9083 &0.0042 &0.0375 &0.9582 &38.6344 &2.7254 &3.6221 &0.8558 \\
LFormer  &0.0124 &0.0277 &0.9602 &50.2647 &0.6210 &0.8092 &0.8988 &0.0253 &0.0541 &0.9227  &38.0189 &2.9513 &3.8192 &0.8493  \\
LFormer-C &0.0053 &0.0272 &0.9676 &50.3799 &0.6149 &0.7990 &0.9006 &0.0049 &0.0380 &0.9572  &38.2575 &2.8623 &3.7609 &0.8506  \\
PanMamba  &0.0134 &0.0277 &0.9592 &50.5325 &0.5927 &0.7778 &0.8998 &0.0152&0.0429&0.9426 &38.1740 &2.8900 &3.7766 &0.8504 	\\
PanMamba-C  &0.0050 &0.0278 &0.9672 &50.6724 &0.5888 &0.7731 &0.9022  &0.0051 &0.0371 &0.9578 &38.2407 &2.8548 &3.7540 &0.8513 \\
\bottomrule
\end{tabular}
\caption{Quantitative results on the full-resolution and reduced-resolution datasets.}
\label{tab_1}
\end{table*}
\section{Experiments}

\subsection{Experimental Settings}

\subsubsection{Datasets.}
Our experimental datasets are captured from the WorldView-3 (WV3) sensor, comprising eight spectral bands, as well as a four-band dataset from the QuickBird (QB) sensor. The QB and WV3 datasets are divided into non-overlapped training, validation, and test sets, respectively.

\subsubsection{Training Details.} All models in paper were trained on a desktop computer equipped with a GTX-4090 GPU. The update of the CLIPPan framework is optimized by the Adam optimizer with a 0.003 learning rate. The batch size and the iteration number are set to 32 and 1000, respectively.

\subsubsection{Metrics.} For reduced-resolution experiment, we adopt four indicators for assessment: the mean peak signal-to-noise ratio (MPSNR), the erreur relative globale adimensionnelle de synthèse (ERGAS) \cite{ref40}, the spectral angle mapper (SAM) \cite{ref41}, and the Q2n index \cite{ref42}, i.e., Q4 for four band and Q8 for eight band datasets. For full-resolution experiments, we use the quality without a reference (QNR) index \cite{ref43} and the related spatial ($D_{\mathrm{s}}$) and spectral ($D_{\mathrm{\lambda}}$) distortion indexes.

\subsection{Effectiveness for Different Baseline Methods}
\subsubsection{Quantitative Comparison.}
To validate the universality of the proposed CLIPPan framework, four representative pansharpening backbones—$\lambda$-PNN \cite{lamdpnn}, PNN \cite{ref17}, ArbRPN \cite{ref36}, LFormer \cite{lformer}, and Pan-Mamba \cite{panmamba} trained with CLIPPan are denoted “Method-C”, while the baseline version employs the conventional loss. For valid and fair comparison, full-resolution and reduced-resolution experiments adopted unsupervised and supervised training approaches, respectively. The traditional SR-D~\cite{ref45} and MTF-GLP-HPM-R~\cite{ref46} are employed solely as comparative methods. Table~\ref{tab_1} reports the quantitative results. Across the board, integrating CLIPPan yields consistent improvements for all backbones. Specifically, compared with baseline methods, ArbRPN-C achieves a 79\% reduction in spectral distortion $D_{\mathrm{\lambda}}$ and enhances QNR by 0.011 on the QB dataset. Similarly, LFormer-C decreases spatial distortion $D_{\mathrm{s}}$ by approximately 30\% and improves QNR by 0.035 on the WV3 dataset. These results conclusively demonstrate that the CLIPPan framework has the capability of transferring the language descriptions in Wald's protocol to pansharpening outcomes. That is to say, even without ground-truth data, the proposed CLIPPan framework still improves spectral and spatial fidelity for the pansharpening task. 

Additionally, all evaluation metrics of reduced-resolution experimental results have been improved compared with baseline methods. It shows that the CLIPPan framework is also effective for supervised pansharpening.


\subsubsection{Qualitative Comparison.}
Figure~\ref{fig_3} shows the qualitative results on the full-resolution QB dataset. Obviously, results generated by the CLIPPan-integrated (``Method-C") methods have more spatial detail information than those produced by baseline methods. The magnified regions in the bottom row further reveal that integrating the CLIPPan framework into the baseline methods enhances texture details, sharpens outlines, and improves contrast. 


To demonstrate the generalization ability of the proposed CLIPPan framework, we conduct experiments on reduced-resolution QB dataset. As shown in Figure~\ref{fig_4}, it is observed that the Method-C group consistently exhibits the closest texture details and spectral fidelity to the reference image. 


\subsection{Ablation Study}

\begin{table}[t]
\centering
\setlength{\tabcolsep}{3pt} 
\begin{tabular}{lcccc}
\toprule
\cmidrule{1-5}
Method  &MPSNR↑ &ERGAS↓ &SAM↓ &Q2n↑\\ 
\cmidrule{1-5}
$\mathcal{L}_{\text{spec}}+\mathcal{L}_{\text{spat}}$ &29.2739 &8.9584 &9.1671 &0.6102	 \\
$\mathcal{L}_{\text{QNR}}$ &32.0266 &5.9536 &6.7052 &0.7013 \\
$\mathcal{L}_{\text{unsup}}$ &32.1946 &5.8776 &6.6616 &0.7100   \\ \hline
$\mathcal{L}_{\text{unsup}}+\mathcal{L}_{\text{ship}}$ &33.5735 &4.5659 &5.9493 &0.7661 \\
$\mathcal{L}_{\text{unsup}}+\mathcal{L}_{\text{d}}$ &32.3696 &5.7517 &6.5471 &0.7366 \\
$\mathcal{L}_{\text{unsup}}+\mathcal{L}_{\text{ship}}+\mathcal{L}_{d}$ &\textbf{34.7191} &\textbf{4.4922} &\textbf{5.5429} &\textbf{0.7986} \\
\bottomrule
\end{tabular}
\caption {Ablation experiment on unsupervised fusion losses. $\mathcal{L}_{\text{unsup}}$ denotes $\mathcal{L}_{\text{spec}}+\mathcal{L}_{\text{spat}}+\mathcal{L}_{\text{QNR}}$.}
\label{tab_2}
\end{table}

\subsubsection{Effects of Different Unsupervised Loss.} We explore how loss in Stage II affects the pansharpening outcomes by progressively adding loss in Eqs. (\ref{eq:low-level-loss}) and (\ref{eq:semantic-loss}), where the quantitative results are presented in Table~\ref{tab_2}.
The  $\mathcal{L}_{\text{spec}}+\mathcal{L}_{\text{spat}}$ combining pixel-wise spectral and spatial reconstruction terms is adopted as the baseline to ensure the spectral fidelity and spatial sharpness of pansharpened results. 
By integrating the QNR-based loss with these spectral and spatial terms, $\mathcal{L}_{\text{unsup}}$ leads to a significant increase in MPSNR and Q2n by 2.92dB and 0.10, respectively. A significant reduction is also achieved in ERGAS and SAM, 3.08 and 2.51, respectively. Similarly, separately adding either pseudo-supervision loss $\mathcal{L}_{\text{ship}}$ or language-guided semantic loss $\mathcal{L}_{d}$ to the unsupervised loss leads to improvements in all metrics. By contrast, incorporating both losses simultaneously into the unsupervised loss significantly improves all metrics. These results confirm that the combination of unsupervised reconstruction, pseudo-label regularization, and semantic alignment via language in our training loss is the best choice.


\begin{table}[t]
\centering
\setlength{\tabcolsep}{3pt} 
\begin{tabular}{lcccc}
\toprule
\cmidrule{1-5}
Method  &MPSNR↑ &ERGAS↓ &SAM↓ &Q2n↑\\ 
\cmidrule{1-5}
w/o fine-tuning &34.3445 &4.7019 &5.5517 &0.7848 \\
$\mathcal{L}_\text{intra}$ &34.5553 &4.5237 &5.5141 &0.7904\\
$\mathcal{L}_\text{intra}+\mathcal{L}_\text{inter}$ &34.6019 &4.5349 &\textbf{5.2694} &0.7930 \\
$\mathcal{L}_\text{intra}+\mathcal{L}_\text{inter}+\mathcal{L}_{1}$ &\textbf{34.7191} &\textbf{4.4922} &5.5429 &\textbf{0.7986} \\
\bottomrule
\end{tabular}
\caption{Ablation experiments on WV3 for adapting CLIP.}
\label{tab_3}
\end{table}

\subsubsection{Effects of CLIP Fine-Tuning Loss.} We verify the effectiveness of each loss term by adding them gradually to the CLIP adaptation stage. Table~\ref{tab_3} summarizes the results of the ablation experiment on the WV3 dataset. We use the CLIP model without any fine-tuning as the baseline. Therefore, incorporating IntraMCL loss $\mathcal{L}_{\text{intra}}$ improves the model performance across all metrics. These indicate that $\mathcal{L}_{\text{intra}}$ loss promotes the CLIP model to recognize remote-sensing image content, providing valuable semantic information for the downstream pansharpening task. Similarly, incrementally adding InterMCL loss $\mathcal{L}_{\text{inter}}$ further enhances model performance, particularly improving MPSNR by 0.26dB and reducing SAM by 0.28. These demonstrate that $\mathcal{L}_{\text{inter}}$ loss enables fine-grained alignment between visual and textual representations, thereby achieving modality-invariant feature learning. Further applying the $\mathcal{L}_{1}$ loss, we observe sustained improvements in all evaluation metrics. These results reflect that the incorporation of IntraMCL, InterMCL, and $\mathcal{L}_{1}$ constraints on CLIP fine-tuning achieves best pansharpening performance.


\begin{table}[t]
\centering
\begin{tabular}{lcccc}
\toprule
\cmidrule{1-5}
Method  &MPSNR↑ &ERGAS↓ &SAM↓ &Q2n↑\\ 
\cmidrule{1-5}
PCA &34.6903 &4.7911 &5.6647 &0.7955\\
RGB &34.4237 &4.6220 &5.6168 &0.7940 \\
GBNIR&34.3282 &4.5038 &5.6804 &0.7957\\
Conv &\textbf{34.7191} &\textbf{4.4922} &\textbf{5.5429} &\textbf{0.7986} \\
\bottomrule
\end{tabular}
\caption{Ablation experiment on feature extraction of MS images in CLIP. PCA: principal component analysis; RGB: direct RGB extraction; GBNIR: Green-Blue-NIR composite; Conv: learnable residual convolution.}
\label{tab_4}
\end{table}

\subsubsection{Effects of Different MS Input Manner.} To ensure compatibility with CLIP’s three-channel input, we evaluated four strategies for compressing the multi-band MS image, as summarized in Table~\ref{tab_4}. Results indicate that directly utilizing RGB or GBNIR channels for extraction struggles to achieve satisfactory pansharpening performance. Although the PCA approach improves the MPSNR metric, it leads to significant degradation in other metrics, particularly with a substantial drop in ERGAS. In contrast, the proposed learnable residual convolution yields the best trade-off across all metrics, confirming that a data-driven channel projection preserves both spectral fidelity and spatial detail.


\begin{table}[t]
\centering
\begin{tabular}{lcccc}
\toprule
\cmidrule{1-5}
Method  &MPSNR↑ &ERGAS↓ &SAM↓ &Q2n↑\\ 
\cmidrule{1-5}
Noise &34.3671 &4.6553 &5.8676 &0.7869\\
I &34.5060 &4.6490 &\textbf{5.4767} &0.7911\\
II &\textbf{34.7662} &4.5178 &5.6075 &0.7968\\
Khan's  &34.6581 &4.5021 &5.5506 &0.7969\\
Wald's  &34.7191 &\textbf{4.4922} &5.5429 &\textbf{0.7986} \\
\bottomrule
\end{tabular}
\caption{Ablation experiment on textual descriptors for HRMS images. ``Noise descriptor” states ``an image independent of the inputs," ``I” denotes “This image is the fusion image of the input image,” and ``II” denotes``a fused product of the MS and PAN images”.}
\label{tab_5}
\end{table}

\subsubsection{Effects of Different Textual Fusion Descriptors.} 
Table~\ref{tab_5} lists results of textual fusion descriptions on final performance. The Noise descriptor and Khan's protocol~\cite{kan} fail to achieve the best effect on any metric. Unfortunately, neither ``I" nor ``II" description approach can effectively optimize both MPSNR and SAM metrics simultaneously. Overall, Wald's method achieved optimal balance in all metrics. These findings demonstrate that precise, protocol-compliant text supervision is crucial for leveraging CLIP's semantic space in unsupervised pansharpening.


\section{Conclusion}
In this paper, we presented CLIPPan, a novel unsupervised pansharpening framework that repurposes CLIP as a language-driven supervisor. By lightweight fine-tuning, CLIP learns to recognize LRMS, PAN, and HRMS types and enforces fusion rules such as Wald’s protocol purely through text. Extensive experiments on real full-resolution imagery show that CLIPPan consistently boosts existing backbones, outperform the SOTA unsupervised framework, and sets new state-of-the-art results without any GT.

\section{Acknowledgments}
This work is supported in part by the National Natural Science Foundation of China under Grant 62101502 and 62301093, the China Postdoctoral Science Foundation (CPSF) under Grant 2022T150596 and 2023M730425, the Postdoctoral Fellowship Program of CPSF under Grant GZC20233336, the Key Research \& Development and Promotion Foundation of He'nan Province under Grant 232102211036, and the Natural Science Foundation of Chongqing under Grant CSTB2024NSCQ-MSX0918, and in part by the Natural Science Foundation of Henan under Grant 252300421876. 


\bibliography{aaai2026}

\end{document}